# High-Entropy Solid Electrolytes Discovery: A Dual-Stage Machine Learning Framework Bridging Atomic Configurations and Ionic Transport Properties

*Xiao Fu [a,b], Jing Xu [a,c], Qifan Yang [a,b], Xuhe Gong [a,d], Jingchen Lian [a,c], Liqi Wang [a,c], Zibin Wang [a,c], Ruijuan Xiao [a,b]\*, Hong Li [a,b]\**

[a] Institute of Physics, Chinese Academy of Sciences, Beijing 100190, China
[b] Center of Materials Science and Optoelectronics Engineering, University of Chinese Academy of Sciences, Beijing 100049, China
[c] School of physical sciences, University of Chinese Academy of Sciences, Beijing 100049, China
[d] School of Materials Science and Engineering, Key Laboratory of Aerospace Materials and Performance (Ministry of Education), Beihang University, Beijing 100191, China
\*E-mail: rjxiao@iphy.ac.cn, hli@iphy.ac.cn

**ABSTRACT**

The rapid development of computational materials science powered by machine learning (ML) is gradually leading to solutions to several previously intractable scientific problems. One of the most prominent is machine learning interatomic potentials (MLIPs), which expedites the study of dynamical methods for large-scale systems. However, a promising field, high-entropy (HE) solid-state electrolytes (SEs) remain constrained by trial-and-error paradigms, lacking systematic computational strategies to address their huge and high-dimensional composition space. In this work, we establish a dual-stage ML framework that combines fine-tuned MLIPs with interpretable feature-property mapping to accelerate the high-entropy SEs discovery. Using $Li_3Zr_2Si_2PO_{12}$ (LZSP) as a prototype, the fine-tuned CHGNet-based relaxation provides atomic structure for each configuration, the structure features - mean squared displacement (SF-MSD) model predicts the ionic transport properties and identifies critical descriptors. The theoretical studies indicate that the framework can satisfy the multiple requirements including computational efficiency, generalization reliability and prediction accuracy. One of the most promising element combinations in the quinary HE-LZSP space containing 4575 compositions is identified with a high ionic conductivity of 4.53 mS/cm as an application example. The framework contains generalizability and extensibility to other SE families.

**INTRODUCTION**

The fourth scientific and technological revolution represents an intertwining of the energy and information revolutions. The energy revolution is characterized by the advent of clean energy technologies, including photovoltaics, lithium batteries, electric vehicles and energy storage, which are reducing reliance on fossil fuels on a substantial scale[1–3]. The information revolution represents a further leap in computer technology, as evidenced by the explosion of artificial intelligence[4]. In the past decades, computational materials science has been developing fast and playing a crucial role in the field of battery materials, particularly in research of electrode reaction process simulation[5–7], electrolyte-electrode interface stability analysis[8–10], and the design of new solid-state electrolytes (SEs)[11–13]. However, the high-entropy (HE) strategy, which has recently received much attention in the field of SE modification and has shown promising potential, still faces significant practical challenges when investigated by traditional computational methods[14]. The HE strategy in SE research introduces a large number of partially occupied sites in the lattice, causing structural distortions. As a result of these distortions, the scale of the bottleneck for cation migration changes. Under the appropriate conditions, these changes can facilitate the formation of an ionic percolation network with a low migration barrier, ultimately leading to high ionic conductivity that surpasses that of conventional low-entropy materials[15,16]. This contributes to the development of all-solid-state batteries (ASSBs) with high power density and performance capability. A similar situation in some partially occupied structures is known as the frustration mechanisms, and Wang et al. run MD for extended time scale and then elucidate how frustration enhances diffusion through the broadening and overlapping of the energy levels of atomistic states[17]. Nevertheless, its rapid development is restrained by two key factors[14]: first, experiments are hard to identify reasonable compositions directly and must rely on repeated trial-and-error; second, computational prediction must contend with a large simulation cell and a vast composition space resulting from the proportion of elemental species and ratio, which are challenging for traditional computational methods to handle, particularly when it is inevitable using molecular dynamics (MD) simulations.

To resolve the issues currently facing in this field, the integration of machine learning (ML) strategies might prove an effective solution. In the field of computational materials science, researchers have developed a variety of different methods to train ML interatomic potentials (MLIPs), including DeePMD[18], NequIP[19], CHGNet[20] and so on. These have been demonstrated to offer enhanced simulation accuracy and efficiency in a range of research efforts[21–23]. Meanwhile, ML models based on structure-property relationship have also attracted considerable attention as a tool for elucidating the causes of superior material properties and accelerating the prediction of target properties[24–26]. For instance, the Sure Independence Screening and Sparsifying Operator (SISSO) model developed by Ouyang et al[27] can construct accurate and explainable prediction models for material behaviors[28–30].

In this study, a comprehensive approach is established for the speedy and stable prediction of the ionic transport properties of high-entropy SEs. As an emerging member of NASICON structured SEs in recent years, $Li_3Zr_2Si_2PO_{12}$ (LZSP) has received extensive attention from researchers, and its ionic transport properties are highly sensitive to bottleneck sizes[31], so it is suitable to be selected as a pristine material for HE SEs to study the structure-property relationship in HE SEs and further adjust its structural framework to enhance the ionic conductivity. Firstly, we construct datasets containing information on multiple elemental interactions and transition states for Li migration and fine-tune the pretrained CHGNet to obtain high-precision MLIPs. Further, structural features and MD simulations for hundreds of structures are carried out, to develop the 'Structural Features to Mean-Squared Displacement' (SF-MSD) model to promptly evaluate the ionic transport properties of materials within the vast HE composition space. Based on this dual-stage framework, we perform a thorough evaluation of a branch of quinary HE LZSPs and discover $Li_{2.625}Zr_{0.25}Hf_{0.1875}Sn_{0.1875}Ti_{0.1875}Nb_{0.1875}Si_2PO_{12}$ (LZHSTNSP) with a three-order-of-magnitude improvement of extrapolated room-temperature ionic conductivity compared to the pristine LZSP, illustrating the reasonableness of this workflow and unveiling what structural features lead to superior properties.

**2. RESULTS AND DISCUSSION**

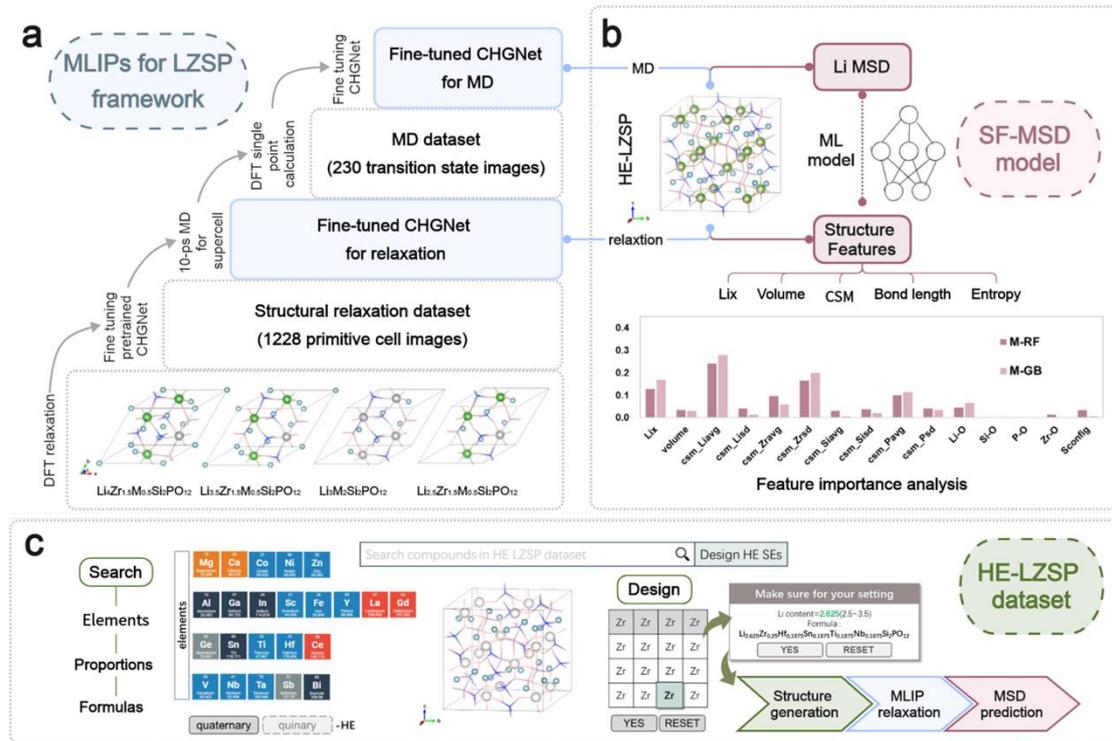

Fig 1. Workflow for (a) preparation of MLIPs for LZSP framework and (b) development of the SF-MSD model. (c) The HE-LZSP dataset, constructed from data and models from this work, facilitates not only the search of transport properties but also the exploration and design of new compositions.

## 2.1. Fine-tuned CHGNet in LZSP Framework

The vast majority of current computational methods for inorganic materials are based on periodic boundary conditions, and because of this, even with the benefit of high-throughput computing and supercomputers, any HE material that is to be simulated by first-principles calculation requires a great deal of computational effort, since the intricate nature of the disordered occupied state necessitates the utilization of a sufficiently expansive supercell within the computational process.

CHGNet is an MLIP that could be suitable to be used to investigate multinary systems through large-scale computational simulations[20]. The CHGNet model, pretrained on the Materials Project trajectory dataset, can achieve DFT-level accuracy through targeted fine-tuning with minimal system-specific data. The stepwise fine-tuning for structural relaxation and MD simulations has been chosen as a solution to the problem of potential energy surface softening and energy distribution[32]. In general, the elements chosen for the design of HE SEs have similar properties as the elements in the original material, and they are also most likely to be used for its doping modification, so that a sufficiently large number of elemental species should be considered in the dataset construction within the elements used for doping. In addition to this, LZSP and $Na_3Zr_2Si_2PO_{12}$ have identical structural frameworks and cation transport paths, from which it can be inferred that the Li content is potentially influential on the stability and transport properties of the structures, and it is necessary to carry out the MLIP training dataset construction and subsequent studies in an appropriate Li content range. In order to guarantee both computational precision and wide scope of the MLIP for HE-LZSPs, the chemical compositions $Li_4Zr_{1.5}M_{0.5}Si_2PO_{12}$ (M = Mg, Ca, Zn, Ni, Co), $Li_{3.5}Zr_{1.5}M_{0.5}Si_2PO_{12}$ (M = Al, In, Ga, Sc, Y, Fe, La, Gd), $Li_3M_2Si_2PO_{12}$ (M = Hf,

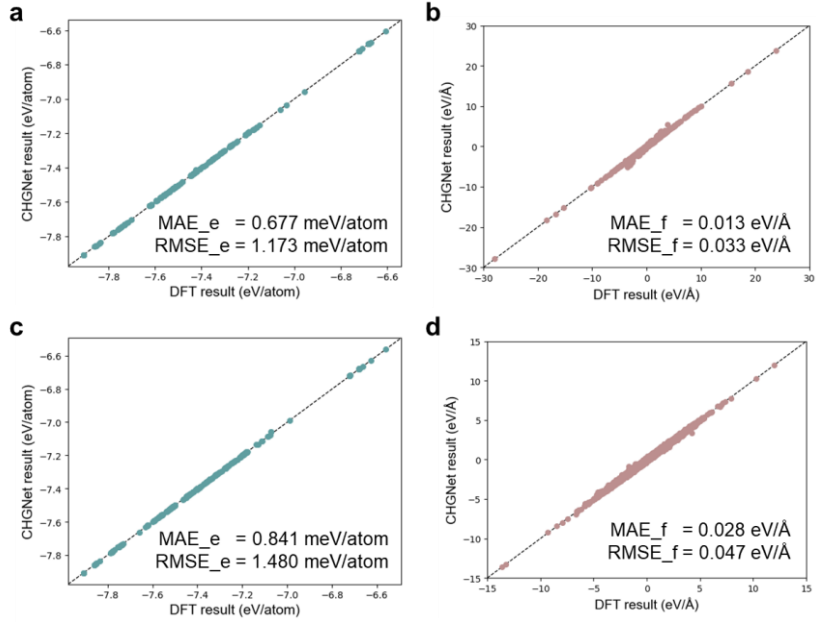

Fig 2. Testing set performance of the fine-tuned CHGNet on predicting energies and forces for (a, b) structural relaxation and (c, d) MD dataset, respectively.

Ge, Sn, Ti, Ce)and $Li_{2.5}Zr_{1.5}M_{0.5}Si_2PO_{12}$ (M = Nb, Ta, Sb, Bi, V) are involved in the dataset. The structural relaxation is conducted in the primitive cells, with up to 50 images for each composition selected for incorporation into the structural relaxation dataset, which ultimately contains 1228 images. The fine-tuned structural relaxation CHGNet model demonstrates high accuracy on the test set as illustrated in Fig 2(a, b), employing this model to calculate 10-ps MD of supercells at 800 K with recording 1000 images for each composition. From these images, 10 structures that contain the most transition state information according to the Li atom positions are selected to make up the MD dataset containing 230 data. Further fine-tuning CHGNet, starting with the relaxation CHGNet model, to align with MD simulation conditions also reaches satisfactory accuracy, as shown in Fig 2(c, d). The inherent robustness of the CHGNet pretrained model as a general-purpose force field is a key factor in the rapid improvement across all evaluation metrics that is enabled by fine-tuning with minimal amount of raw data, achieving accuracy comparable to DFT calculations.

    The generalizability of MLIPs is a critical metric for evaluating their performance, particularly when predicting structurally similar but compositionally distinct configurations compared to training data. When extending the application of the fine-tuned CHGNet model from single-element-doped structures to more intricate HE-LZSP family as discussed in subsequent sections, a rigorous evaluation of its generalization performance is imperative. Considering that the following exploration involves variations in lithium content, elemental composition and stoichiometry, we perform structural relaxation and accuracy testing on randomly selected configurations from Table 1 using both CHGNet and DFT calculations, and the key metrics including computational time, final energies, and unit cell volumes are compared systematically between two methods. The fine-tuned CHGNet demonstrates the effectiveness as an alternative to DFT for structural relaxation, achieving cell volume error less than 1% and root sum squared distance (RSSD) error of lattice sites under 1Å in most cases. Notably, these structural optimization tests are conducted in primitive cells, with up to four different elements occupying the Zr sites. The simulations of supercells with more than four Zr sites are prohibitively expensive for DFT relaxation calculations. Meanwhile, whether slightly

Table 1. Tests of fine-tuned CHGNet for structural relaxation.

| Composition | DFT relaxation | | | CHGNet relaxation | | | ΔV% | RSSD /Å |
|---|---|---|---|---|---|---|---|---|
| | Time/ | E/eV | V/Å³ | Time/s | E/eV | V/Å³ | | |
| $Li_{3.5}La_{0.5}Zr_{1.5}Si_2PO_{12}$ | 18736 | -308.259 | 522.303 | 53 | -308.237 | 523.313 | 0.19 | 0.309 |
| $Li_3SnZrSi_2PO_{12}$ | 6860 | -287.755 | 504.650 | 29 | -287.410 | 502.557 | 0.41 | 0.230 |
| $Li_{2.5}Bi_{0.5}Zr_{1.5}Si_2PO_{12}$ | 30919 | -290.806 | 518.689 | 94 | -291.023 | 519.566 | 0.17 | 0.186 |
| $Li_4Ga_{0.5}Y_{0.5}ZrSi_2PO_{12}$ | 17841 | -302.775 | 539.621 | 49 | -303.584 | 535.673 | 0.73 | 0.897 |
| $Li_3NbZn_{0.5}Zr_{0.5}Si_2PO_{12}$ | 14318 | -292.990 | 500.090 | 43 | -294.174 | 498.545 | 0.31 | 0.545 |
| $Li_{2.5}Sb_{0.5}Ti_{0.5}ZrSi_2PO_{12}$ | 6848 | -290.658 | 498.409 | 22 | -290.865 | 498.071 | 0.07 | 0.133 |
| $Li_4Fe_{0.5}Al_{0.5}Ge_{0.5}Zr_{0.5}Si_2PO_{12}$ | 44082 | -291.947 | 492.104 | 692 | -292.253 | 519.138 | 5.49 | 4.680 |
| $Li_3V_{0.5}In_{0.5}Hf_{0.5}Zr_{0.5}Si_2PO_{12}$ | 10746 | -291.587 | 496.506 | 33 | -292.461 | 497.253 | 0.15 | 0.392 |
| $Li_{2.5}Nb_{0.5}Bi_{0.5}Fe_{0.5}Zr_{0.5}Si_2PO_{12}$ | 32692 | -280.324 | 502.286 | 104 | -282.045 | 499.451 | 0.56 | 2.434 |

higher relaxation errors observed in a few structures make a critical impact on final results are discussed in detail in the Supplementary text 2. In addition, a generalization test of fine-tuned CHGNet for MD is also performed to compare with *Ab initio* molecular dynamics (AIMD) using the quaternary structure $Li_{3.5}Zr_{0.5}Hf_{0.5}Al_{0.5}Ge_{0.5}Si_2PO_{12}$. To reduce the statistical errors in AIMD data analysis, we utilized the 20~80 ps time interval from the 100 ps trajectory to calculate the MSD-Δt curves[33]. The statistically obtained bond length distribution and MSD curves match significantly as displayed in Fig 3. Collectively, the outcomes prove that the fine-tuned CHGNet achieves an optimal balance between computational accuracy and efficiency, establishing its reliability for both the structural relaxation and ionic migration analysis in HE-LZSP systems.

**2.2. Quaternary SF-MSD Model**

The fine-tuned CHGNet model can accelerate the relaxation and simulation to evaluate ionic transport properties of HE-LZSPs, but in the face of numerous configurations in HE systems, it is impossible to exhaust the entire chemical space. Therefore, the ML, structure features - mean squared displacement (SF-MSD), model is established on the basis of the fine-tuned CHGNet calculation results. One of the most important aspects of ML model building is the selection of features and datasets, in the next step we have taken quaternary structures as an example to try the initial construction of the SF-MSD model and analyze in detail the feasibility of the whole workflow of HE-LZSP transport property prediction.

With the wide application of ML models in material science, various atomic structure descriptors have been developed as data inputs, for example atom-centered symmetry functions (ACSF)[34] and smooth overlap of atomic positions (SOAP)[35] descriptors. However, all the structures investigated in this work are based on the same LZSP framework, hence to enhance training efficiency of the SF-MSD model, we strategically selected 17 representative structural parameters as input feature values, including the lithium content, the cell volume and the X-O bond lengths and continuous symmetry metrics (CSM) [13,36] values of $XO_n$ polyhedra that characterize the oxygen frameworks of the relaxed structures. Subsequently, feature importance analyses are conducted based on these parameters to investigate the transport mechanisms in HE-LZSPs.

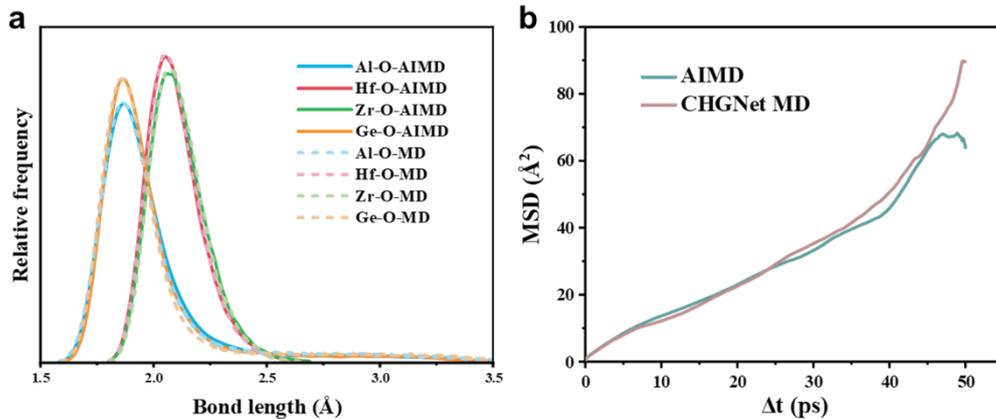

Fig 3. The generalization performance of the fine-tuned CHGNet for MD. (a) Bond length distribution and (b) MSD curves of quaternary structure $Li_{3.5}Zr_{0.5}Hf_{0.5}Al_{0.5}Ge_{0.5}Si_2PO_{12}$.

In order to establish a test case, we initially examine the special configuration of the four elements occupying Zr sites with 1:1:1:1 ratio. According to the elemental combinations covered in our existing dataset, we generated 735 compositions with lithium content ranging from 2.5 to 3.5. This dataset was partitioned, allocating a small subset as the training set for the SF-MSD model and the majority as the predicted targets. Our fine-tuned CHGNet model is used to perform structural relaxation on all the configurations and conduct MD simulations at 1000 K for nearly 200 compositions to obtain the target MSD values. It is worth noting that the majority of the MSD values obtained from MD simulations of these structures are distributed in the lower range as shown in Fig S1(a), which explains the sluggish progress of HE SEs research based on the conventional trial-and-error approaches. The systematic analysis indicates that the structural regulation through framework distortions is not a simple combination of several desirable dopant elements. Furthermore, excessive concentration of training samples in the lower ranges may compromise predictive performance for other regimes, and sparse target value distributions of the outliers could lead to problems of model stability. To address these challenges in predicting HE-LZSPs with elevated MSD values, we implemented the dataset refinement by selective removal of the most proximate target values, resulting in a relatively more balanced MSD distribution. After comparing typical ML approaches for nonlinear regression tasks, we found that the gradient boosting (GB) and random forest (RF) model can achieve greater accuracy, and the SISSO model performed well in structure-property relationship predictions[28–30], thus we chose these three methodologies as candidate models for systematic comparison, as illustrated in Fig 4. In the figure there are also samples with MSD > 200 $Å^2$ labeled with stars, supplementing regular training and testing error. Although these values show noticeable absolute deviations, the SF-MSD model can distinguish them from structures with inferior transport properties through trend analysis. Notably, even though the SISSO model shows the smallest maximum absolute error (MaxAE) on the training set, its error on the testing set escalates significantly, which indicates that the model is unsuitable for structures beyond training configurations. In addition to discrepancies in model performance, obvious differences in feature importance distribution also emerges between SISSO and the GB/RF models, as demonstrated in Fig 4(d).

The comprehensive evaluation of the SF-MSD model requires taking into account the workflow validity on the integration of fine-tuned CHGNet and SF-MSD models, as well as the prediction accuracy for target MSD values. The workflow comprises three key steps:

**Step 1.** Determine the composition formulas and generate randomized atomic configurations.
**Step 2.** Relax the structure with fine-tuned CHGNet potentials.
**Step 3.** Extract the relaxed structural features for SF-MSD model to conduct MSD prediction.

To ensure representative sampling of the random generation for the disordered atomic arrangement in Step 1, we took $Li_{2.5}Zr_{0.5}Ta_{0.5}Hf_{0.5}Ti_{0.5}Si_2PO_{12}$ as our benchmark study with the following process: Within the 16 Zr-site supercell, 1000 different atomic configurations are randomly generated through stochastic site occupancy. Structural relaxation using fine-tuned CHGNet revealed the maximum energy interval of 9.45 meV/atom and the maximum volume interval of 40.04 Å$^3$ (~2%) across all the configurations, which indicates that a large number of random structures are close to each other, collectively representing a subset of the structural features. Subsequently, the structural features are extracted to carry out MSD predictions with SF-MSD model, and the results are given in Fig 5, which demonstrates enhanced predictive stability in GB and RF models evidenced by their smaller standard deviation (SD) compared to SISSO. This robustness persists despite the atomic arrangement variations and local minima dependencies in relaxation algorithms, ensuring that minor structural perturbations do not induce significant MSD prediction discrepancies in this optimized ML frameworks.

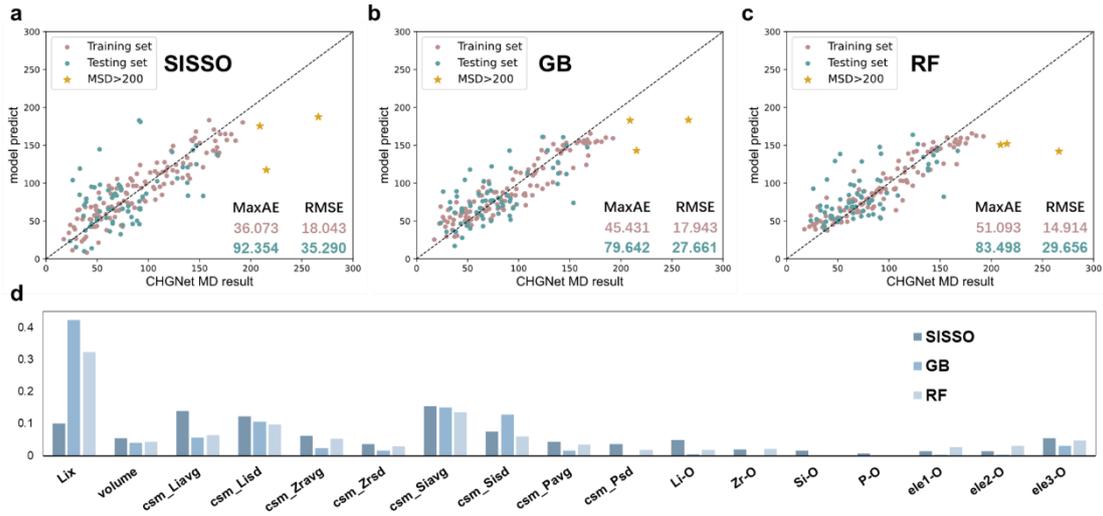

Fig 4. Comparation of different type of SF-MSD model for quaternary structures about (a-c) max absolute error (MaxAE), root mean square error (RMSE), and (d)feature importance, here Lix represents lithium content, csm_Xavg represents average CSM value of atoms at X sites, csm_Xsd represents standard deviation of CSM values of atoms at X sites and X-O represents bond length between X atoms and O atoms read from first nearest neighbor peak of radial distribution function (RDF), respectively.

Through the above study, we confirm the validity of the developed MSD prediction workflow for HE-LZSPs. Besides, we identified that the change of X-O bond length exhibits negligible impact on the MSD prediction, enabling appropriate simplification in the subsequent modeling stages. This simplification proves particularly advantageous for extending the model to the multinary system applications, as not only reducing the input feature dimensionality, but also beneficial to expand the

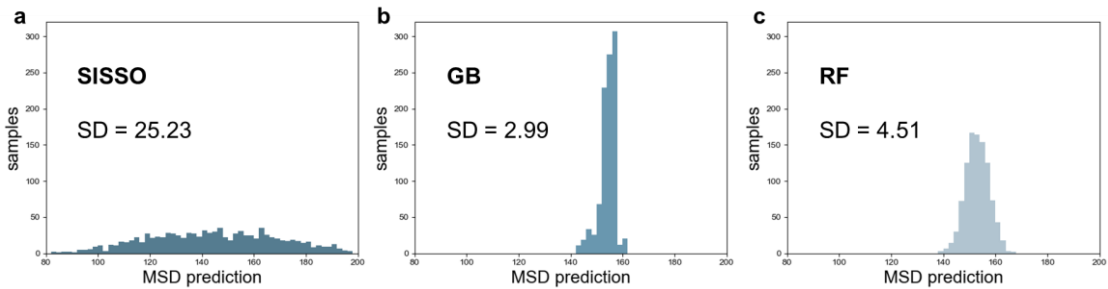

Fig 5. The MSD prediction distributions for (a) SISSO, (b) GB and (c) RF models of 1000 random structures in formula $Li_{2.5}Zr_{0.5}Ta_{0.5}Hf_{0.5}Ti_{0.5}Si_2PO_{12}$

applicability to broader HE composition spaces.

**2.3. Multinary SF-MSD Model**

After establishing theoretical validity for evaluating ionic migration properties in quaternary HE SEs through structural feature analysis using SF-MSD model, we further extended the idea for the prediction in multinary systems to overcome the inherent limitations of models trained on specific composition ratios. According to the feature importance distribution results in Fig. 4(d), adjusting the quaternary model to different composition involves just minor parameter changes, which indicates strong potential for successful generalization of the model to multinary systems. Besides, it can be observed that volume has a discernible impact on the predicted values. Therefore, the volume is decomposed into more specific features, the unit cell parameters, which are directly used for subsequent model training. Compared to the quaternary case, several critical innovations are introduced in the multinary SF-MSD model. The first one is the incorporation of the ideal configuration entropy ($S_{config}$) as the additional feature parameter, which is not considered as an input in the quaternary system because all structures there have the same $S_{config}$ value. Secondly, more combinatorial compositions are created. Three different proportions are selected in each of the ternary to octonary structures, corresponding to the three largest $S_{config}$, as shown in Table S2, and 15 random structures are constructed for each proportion under the condition of controlling the occurrence probability of each element in the dataset, yielding a total of 270 configurations, which are simulated with fine-tuned CHGNet to obtain the calculated values of the MSD, as the multinary dataset. Following the data balancing strategy established for quaternary system, we selected 40 quaternary data and 148 samples from the 270 multinary data to construct the training set for multinary model. The rest 122 samples are taken as multinary testing set, and the quaternary dataset in Section 2.2 is used as the quaternary testing set for additional validation. The multinary SF-MSD model is trained and quantified in Table 2. As shown in Fig S2, the multinary SF-MSD model outperforms the quaternary counterpart, with more balanced distribution of feature importance, which facilitates the accurate grasping of the effective information from the structural feature parameters. Moreover, the cross-validation is conducted on the existing data and the details are in supplementary text 3. The results confirm that the current model maintains strong predictive performance, as effective dataset generation strikes an optimal balance between computational cost and model accuracy. Additionally, the GB model exhibits better robustness than the RF model, demonstrating higher tolerance to unavoidable noises and perturbations in the dataset.

After successfully building the multinary SF-MSD model with sufficient prediction accuracy and generalizability, we deploy this framework to investigate the unexplored quinary HE-LZSPs. In the quinary HE-LZSP space, 4575 distinct 5-element configurations of Type 5-1 in Table S2 are generated by combinatorial element permutations through all the discussed elements. These configurations are categorized into three regions of high, medium and low according to the predicted MSD values (MSD>120 for high, 40<MSD<120 for medium, MSD<40 for low). In each region 10 structures are randomly selected to carry out the fine-tuned CHGNet-based MD simulations, and the validation results are presented in Fig 6(a). Most of the points fall into the acceptable error margin (gray shaded area) with absolute MSD errors less than 30 Å$^2$. Those with larger errors are mostly from the high MSD region, which is due to the fact that the samples in the high MSD range of the original dataset are relatively sparse, corresponding to the truth that only a few compositions in the actual HE-LZSP space are able to achieve apparent high ionic conductivities. In spite of high absolute MSD errors, the relative errors, generalizable between MSD, diffusion coefficient and ionic conductivity, are stable in this region as shown in Fig 6(b), and this indicates that the transport properties predicted by the SF-MSD model do not exhibit order-of-magnitude differences. However, it is worth highlighting that this approach successfully identified an HE-LZSP, $Li_{2.625}Zr_{0.25}Hf_{0.1875}Sn_{0.1875}Ti_{0.1875}Nb_{0.1875}Si_2PO_{12}$ (LZHSTNSP), with remarkable room-temperature ionic conductivity enhancement by three orders of magnitude compared to pristine LZSP as shown in Fig 6(c). With the proper HE-strategy adopted in LZSP, the ionic conductivity at room temperature, extrapolated from the Arrhenius curve, increased to 4.53 mS/cm. Simultaneously, the migration energy barrier was reduced from 0.46 to 0.24 eV in comparison to the pristine LZSP. It should be noted that the reference data for pristine LZSP is also obtained from the fine-tuned CHGNet model, and the consistent results derived from the AIMD simulations can be found in Fig. S4 of the supplementary material.

Table 2. Comparison of RMSE($Å^2$) for models on training set, multinary and quaternary testing set.

| Dataset | Quaternary model | | Multinary model | |
|---|---|---|---|---|
| | GB | RF | GB | RF |
| Training set | 17.943 | 14.914 | 11.986 | 11.136 |
| Testing set (multinary) | × | × | 20.317 | 20.108 |
| Testing set (quaternary) | 27.661 | 29.656 | 23.856 | 24.792 |

Additionally, based on the 4575 structures, the Shapley Additive Explanations (SHAP) analysis was carried out to reveal critical structure-property relationships governing ionic transport in HE-LZSPs, and the results were illustrated in Fig S5. The most important features are the introduction of Li vacancies and its larger distortional oxygen coordination environment, which is in agreement with the idea of corner-sharing framework solid-state electrolyte screening discussed by Jun et al[13]. Secondary but significant contributors are from the local disorder configurations of $XO_6$ octahedra and $PO_4$ tetrahedron, induced by the replacement element of the Zr sites, causes impacts on MSD. As the transport pathways are oriented in a three-dimensional distribution, the overall high degree of anisotropy in the supercell parameters can be detrimental to the connectivity associated with the transport pathways. Taking the promising candidate LZHSTNSP predicted by multinary GB model as an example, Fig S6 shows the waterfall plot for its structural features. The excellent transport property in this structure is mainly due to the uniform Zr-site octahedral distortions and the reasonable Li vacancy concentration, all above make LZHSTNSP a promising fast ionic conductor.

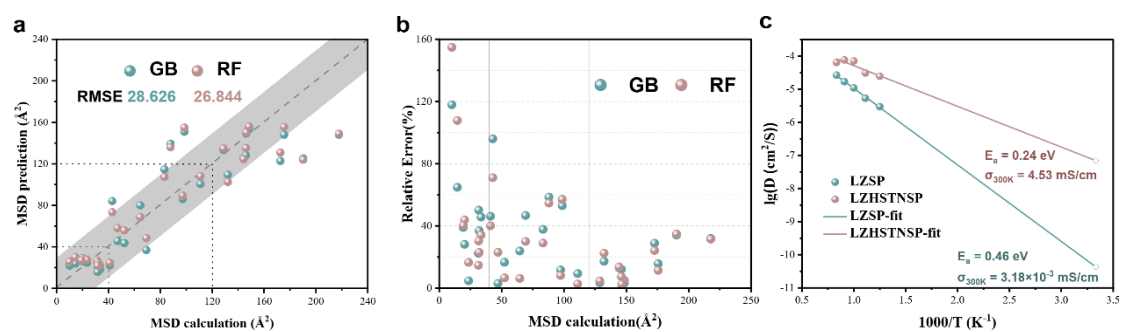

Fig 6. (a) MSD values and (b) relative errors of the SF-MSD model for predicting quinary HE-LZSPs, (c) Arrhenius curves of one of the highest predicted MSD in quinary HE-LZSPs, $Li_{2.625}Zr_{0.25}Hf_{0.1875}Sn_{0.1875}Ti_{0.1875}Nb_{0.1875}Si_2PO_{12}$ (LZHSTNSP), in comparison with the pristine LZSP.

**2.4. HE-LZSP Data Resource Utilization**

Our computational workflow has generated an extensive HE-LZSP dataset, including about 500 different HE-LZSPs with fine-tuned CHGNet-based MD simulations, and more than 5000 HE-LZSPs relaxed structures and corresponding predicted MSD values with SF-MSD model. The great majority of these data belongs to quaternary and quinary HE-LZSPs, corresponding to the most actively explored compositional regime in high-entropy SEs research. Researchers can exploit this resource directly to search for HE-LZSPs through elements, proportions and formulas, and get information of the specified composition, which contains the configuration entropy, the structure parameters/file, the predicted MSD value, and its location in the distribution of all MSD values in the dataset. In this manner, the researcher can select materials suitable for use as fast ionic conductors for specific applications. In further profound studies, researchers can employ the fine-tuned CHGNet interatomic potential and SF-MSD model developed in this work to perform structural optimization and MSD prediction for novel HE SEs compositions beyond existing datasets, enabling rapid assessment of the application potential as fast ion conductors, which is the same process as the one described in Section 2.2. The time consumption of this whole process is in the order of minutes, which is very favorable for both directed design and extensible exploration of HE-LZSP family.

**3. CONCLUSION**

To sum up, we establish a dual-stage ML-based framework that synergistically combines fine-tuned interatomic potentials with interpretable property predictors to accelerate the discovery of HE SEs. Fine-tuned CHGNet achieves DFT-level accuracy while maintaining computational efficiency in the LZSP framework, and the SF-MSD model is constructed to realize the assessment of the transport properties enabling high-dimensional structural feature analysis, thus the combination workflow of the two models realizes the high-throughput property prediction in the HE-LZSP space and reduces the evaluation time for individual compositions from days to minutes. Besides, in quinary HE-LZSPs, the capability of the method to distinguish between structures with varying ionic transport ability has been verified, and a promising HE SE with the composition of $Li_{2.625}Zr_{0.25}Hf_{0.1875}Sn_{0.1875}Ti_{0.1875}Nb_{0.1875}Si_2PO_{12}$ has been screened in our work. Additionally, the SF-MSD model unveils the origin of its excellent transport properties, which is helpful for further directed design and development of HE SEs.

This dual-stage ML-based framework could be expanded to other material families, such as HE garnet and argyrodite SEs, even to other property prediction of HE cathode materials. The comprehensive regime of applicability in computational materials science demonstrates prominent potential for extension beyond HE materials to encompass families of materials sharing a similar structural framework, covering a wide range of chemical compositions. Its versatile applicability offers a unified platform for entropy-driven material design and accelerates discovery across structural families with shared crystallographic framework even with multinary compositions, providing systematic exploration strategy for complex functional materials.

**4. METHODOLOGY**

**4.1. DFT calculations**

Initial structure of pristine LZSP was adopted from ref[31], and the supercell was built by a transformation matrix of $[1, 1, \bar{1}; \bar{1}, 1, 1; 1, \bar{1}, 1]$ acting on the primitive cell. To fine-tune CHGNet for structural relaxation and MD successively, all computational accuracy and setups were consistent in primitive cells and supercells, and MD dataset was generated by single point calculations (NSW=1) of selected images. All DFT calculations with spin on were performed with $\Gamma$-centered $k$-mesh density of one point per 0.1 Å$^{-3}$ using the Vienna *Ab initio* Simulation Program (VASP)[37] within the projector augmented-wave (PAW)[38] approach. The generalized gradient approximation (GGA) [39]was used with the Perdew-Burke-Ernzerhof (PBE)[40] and both ions and cells were relaxed in the optimization with the energy and force convergence criteria of $10^{-5}$ eV and 0.01 eV/Å. For elements that might contain magnetic properties, all parameter settings were the same as the Pymatgen MPRelaxSet which was used in the Materials Project DFT calculations[41,42].

**4.2. MD simulations**

All NVT-Berendsen-MD simulations with fixed lattice were performed using the Atomic simulation environment (ASE) software package[43], the time step was set to 1 fs, and all simulations were set at 1000 K with the exception of the final MD simulation that was used to fit the Arrhenius curve. The simulations performing to compare AIMD and CHGNet MDs lasted 50 ps, while the MD simulations performing to generate the dataset for the SF-MSD model lasted 100 ps. Li MSD was determined as the average of the squared displacement of Li ions at each time step, and to ensure sufficiently reasonable statistics, all the MSD values in the dataset were obtained by taking an average of 20~80 ps over the MSD-Δt curve lasting for 100 ps, which are close to the MSD values obtained at the time interval of 50 ps. By obtaining equivalent values in regions with better linearity during longer simulation times, this method aims to minimize statistical errors in the general slope fitting process as much as possible.

**4.3. Fine-tuning CHGNet**

Based on pretrained CHGNet, the model for structural relaxation and MD were both trained with energy, force, stress and magnetic moment under the mean squared error (MSE) loss criterion and a learning rate of $10^{-3}$ with the RAdam optimizer. The setting for train-validation-test ratio was 6:2:2 with batch size of 50 in structural relaxation model and 10 in MD model, respectively. The model checkpoint of best validation mean absolute error (MAE) was collected for test set predictions, while the final model of 500 epochs was used as a calculator. Other details can be found in Supplementary Material.

**4.4. Training the SF-MSD model**

To construct the structural feature dataset, vasppy and pymatgen software packages[42] were used to calculate the RDF and CSM of the relaxed structures, respectively. Because the RDF was read from only one structure, the curve was smeared with a gaussian kernel of sigma = 0.5. In SISSO model training the following settings were maintained: feature construction operators in '(+)(-)(*)(/)(exp)(exp-)(^-1)(^2)(log)(sin)', 10 of features in each of the SIS-selected subspace, and others as default in SISSO guide input, while the effects of the settings of dimension of the descriptor, maximal feature complexity and the metric for model selection in regression on the prediction results were mainly tested, and the final settings of these three parameters were 7, 4 and 'MaxAE' in that order. At the same time, GB and RF model constructions were executed in scikit-learn procedure package[44]. The Shapley Additive Explanations (SHAP)[45] python package, based on the game theoretic approach, was adopted to dissect the relationship between structural features and ionic transport properties from GB and RF model.

# SUPPLEMENTARY INFORMATION

# High-Entropy Solid Electrolytes Discovery: A Dual-Stage Machine Learning Framework Bridging Atomic Configurations and Ionic Transport Properties

*Xiao Fu [a,b], Jing Xu [a,c], Qifan Yang [a,b], Xuhe Gong [a,d], Jingchen Lian [a,c], Liqi Wang [a,c], Zibin Wang [a,c],*

*Ruijuan Xiao [a,b]\*, Hong Li [a,b]\**

[a] Institute of Physics, Chinese Academy of Sciences, Beijing 100190, China

[b] Center of Materials Science and Optoelectronics Engineering, University of Chinese Academy of Sciences, Beijing 100049, China

[c] School of physical sciences, University of Chinese Academy of Sciences, Beijing 100049, China

[d] School of Materials Science and Engineering, Key Laboratory of Aerospace Materials and Performance (Ministry of Education), Beihang University, Beijing 100191, China

**Corresponding Author**

*E-mail: rjxiao@iphy.ac.cn, hli@iphy.ac.cn

## Supplementary text 1

**Details for CHGNet fine-tuning**

To get more information about elemental interactions and transition states for Li migration from limited DFT data, the following approaches are taken to construct the structural relaxation and MD dataset and complete the CHGNet fine-tuning:

1. Elements are selected by referring research [[1]] and [[2]], and the pristine LZSP structure is adopted from reference[[3]]

2. For structural relaxation dataset, 50 images for each composition consist of the first 10 consecutive high-energy images and follow-by 40 random images from the DFT relaxation ionic steps. This approach produces a dataset with controlled size that better accommodates the large structural changes during the relaxation calculations.

3. For MD dataset, it is important to find the structure of Li in the transition state from a range of structures, and this information is valuable for subsequent simulations of the ionic migration process. In order to find the desired structures, the closest distance from each Li atom for all dynamic images to Li sites in the initial static structure is calculated, and when the distance is greater than 2.5 Å, this Li is considered to be in the transition state, and the structures are sorted according to the number of Li in the transition state, and the top 10 structures containing the most Li atoms in the transition state are added to the MD dataset.

4. It is important to check whether high-energy images are abnormal structures in both of two datasets. If structures of excessively high energies are entered, this can lead to poor model performance or unstable iterations while fine-tuning CHGNet.

## Supplementary text 2

**Discussion of high-error structures in fine-tuned CHGNet relaxation test**

Table 1 shows that the fine-tuned CHGNet demonstrates structural relaxation reproducibility comparable to DFT calculations for most configurations. However, notable exception exists in $Li_4Fe_{0.5}Al_{0.5}Ge_{0.5}Zr_{0.5}Si_2PO_{12}$ as shown in Table 1. We observe volume deviations up to 5.49% and root sum squared distance error up to 4.680 Å. Though, in terms of the accuracy of the CHGNet model, the force errors obtained from the validation on the relaxation path, MAE_f =0.058 eV/Å and RMSE_f = 0.078 eV/Å, are slightly higher than those on its testing set, and such force errors are still tolerable on structures that do not appear in the training dataset. Furthermore, based on these two different structures, CHGNet MD simulations were performed, and the MSD values are obtained by the same method as samples in SF-MSD model dataset. Besides, the SF-MSD model predictions on these two structures were listed in Table S1 together with simulated MSD values. Regardless of the relaxation calculators used to obtain these two structures, the MSD values obtained by the same method are very close. This indicates that the difference between the relaxation of the fine-tuned CHGNet and DFT is not significant enough to affect the subsequent estimation of ionic transport properties. Therefore, it can be considered that this error would not alter the main characteristics of the SF-MSD model.

Table S1. MSD values obtained by different methods based on two configurations of $Li_4Fe_{0.5}Al_{0.5}Ge_{0.5}Zr_{0.5}Si_2PO_{12}$ relaxed by DFT and fine-tuned CHGNet methods.

| Relaxation calculator | Fine-tuned CHGNet MD calculation($Å^2$) | Multinary SF-MSD (RF) prediction($Å^2$) | Multinary SF-MSD (GB) prediction($Å^2$) |
|---|---|---|---|
| DFT | 107.32 | 61.99 | 71.93 |
| Fine-tuned CHGNet | 108.25 | 62.37 | 65.09 |

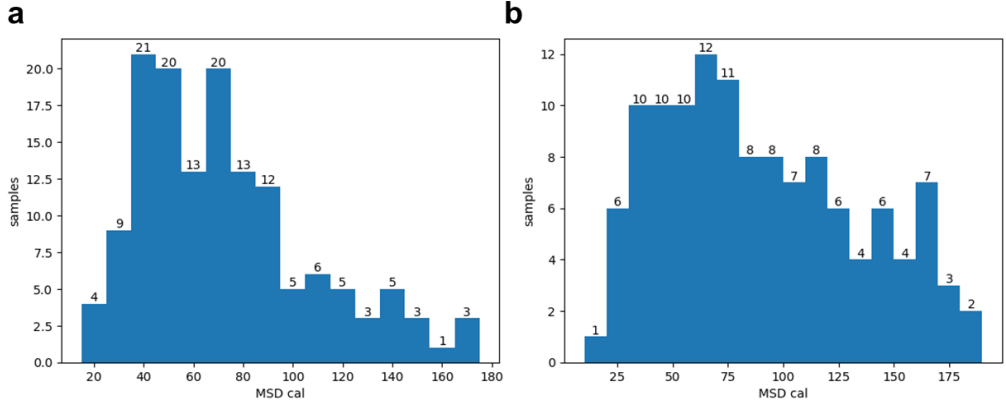

Fig S7. The MSD distribution profiles of (a)randomly chosen samples and (b) data-cleaned samples for quaternary SF-MSD model.

Table S2. Number of sites and $S_{config}$ of different proportion types of compositions in multinary dataset. Here $S_{config}=-R\left(\sum_{i=1}^{N} x_i \ln x_i\right)$, in which $N$ is the number of elemental species, $x_i$ is the mole fraction of component $i$, and R is the universal gas constant (R = 8.314 J K$^{-1}$ mol$^{-1}$), and compositions with $S_{config} \geq 1.5$ R are considered HE, whereas those with 1.5 R > $S_{config} \geq 1$ R and $S_{config} < 1$ R are considered to be medium- and low-entropy systems, respectively. [[4]]

| type | Number of sites (Zr) | | | | | | | | $S_{config}/R$ |
|---|---|---|---|---|---|---|---|---|---|
| | ele0=Zr | ele1 | ele2 | ele3 | ele4 | ele5 | ele6 | ele7 | |
| 3-1 | 6 | 5 | 5 | | | | | | 1.095 |
| 3-2 | 6 | 6 | 4 | | | | | | 1.082 |
| 3-3 | 7 | 5 | 4 | | | | | | 1.072 |
| 4-1 | 4 | 4 | 4 | 4 | | | | | 1.386 |
| 4-2 | 5 | 4 | 4 | 3 | | | | | 1.371 |
| 4-3 | 5 | 5 | 3 | 3 | | | | | 1.355 |
| 5-1 | 4 | 3 | 3 | 3 | 3 | | | | 1.602 |
| 5-2 | 4 | 4 | 3 | 3 | 2 | | | | 1.581 |
| 5-3 | 5 | 3 | 3 | 3 | 2 | | | | 1.565 |
| 6-1 | 3 | 3 | 3 | 3 | 2 | 2 | | | 1.775 |
| 6-2 | 4 | 2 | 2 | 2 | 3 | 3 | | | 1.754 |
| 6-3 | 4 | 4 | 2 | 2 | 2 | 2 | | | 1.733 |
| 7-1 | 3 | 3 | 2 | 2 | 2 | 2 | 2 | | 1.927 |
| 7-2 | 4 | 2 | 2 | 2 | 2 | 2 | 2 | | 1.906 |
| 7-3 | 4 | 3 | 2 | 2 | 2 | 2 | 1 | | 1.873 |
| 8-1 | 2 | 2 | 2 | 2 | 2 | 2 | 2 | 2 | 2.079 |
| 8-2 | 3 | 2 | 2 | 2 | 2 | 2 | 2 | 1 | 2.047 |
| 8-3 | 3 | 3 | 2 | 2 | 2 | 2 | 1 | 1 | 2.014 |

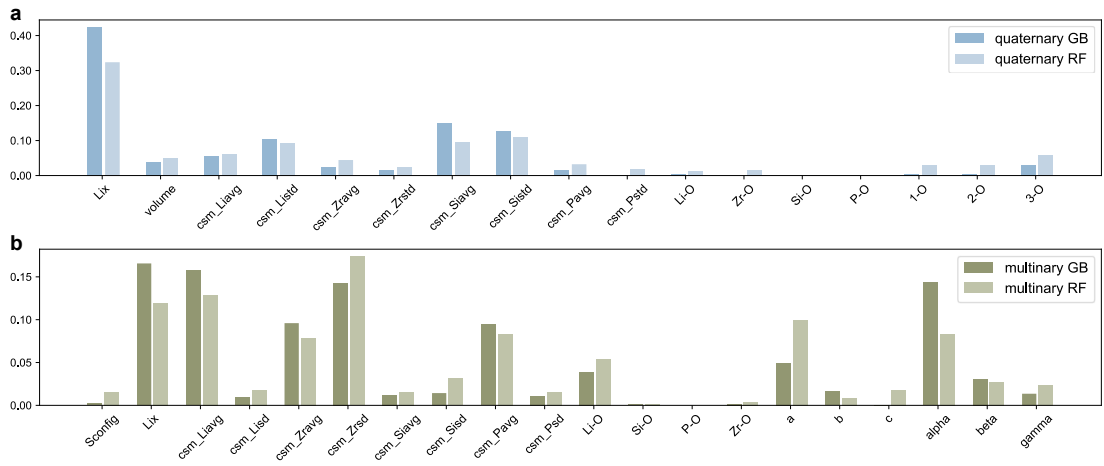

Fig S8. Comparative analysis of feature importance distributions among (a) quaternary model and (b) multinary model.

# Supplementary text 3

**Cross-validation on the combined dataset**

In this work, we expect to achieve a more accurate and faster assessment of the ion transport properties of HE SEs with as little computational effort as possible. To demonstrate that the multinary model in this paper meets the requirements, the following cross-validation was performed with an extended training set. To get a dataset which is the largest in this study for the cross-validation experiment, the quaternary dataset of 172 structures and the multinary dataset of 270 structures are combined, and the testing set is 30 structures in 5-1 type structures as mentioned in Fig. 6 of the main text. Then, the parameters are optimized by randomized search with 100 iterations to obtain models with better predictive performance. It comes out that in the testing set the RMSE of the best GB model is 28.730 Å$^2$ and the best RF model is 26.131 Å$^2$, which is not a significant improvement over the previous model. Therefore, it can be assumed that the models trained using the 188 structures already contain sufficient relationships between structural features and transport properties.

In addition, to compare the robustness of GB and RF models, 1000 different disordered arrangements are tested as input perturbations on the best models in cross-validation. As shown in Fig. S3, the prediction distributions of the GB models are more concentrated than those of the RF models. Furthermore, the best RF model is shifted due to more noise in the combined dataset, while the best GB model is still robust around the same distribution center. Since then, it can be concluded that the GB model has a better robustness than the RF model and can do a better job in SF-MSD prediction work.

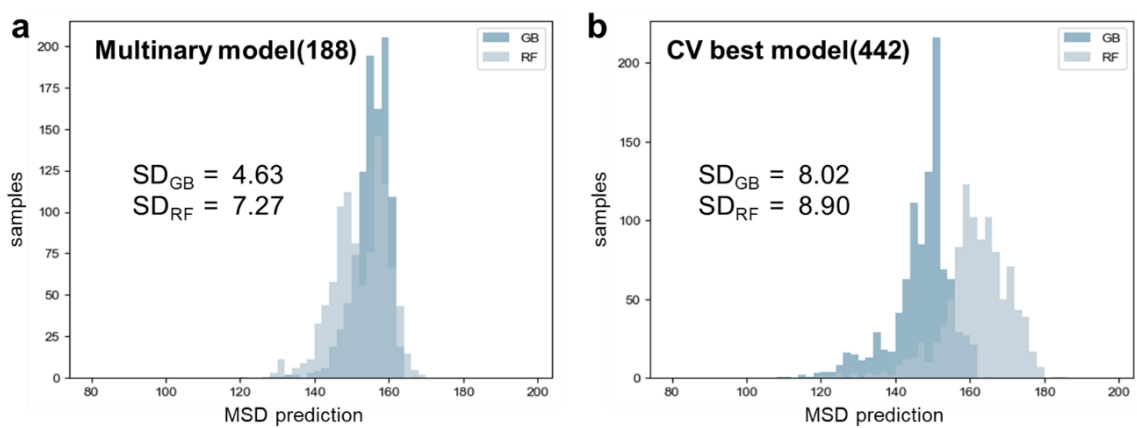

Fig S9 The MSD prediction distributions for (a)the multinary models in the main text and (b)the best models in cross-validation of 1000 random structures in formula Li$_{2.5}$Zr$_{0.5}$Ta$_{0.5}$Hf$_{0.5}$Ti$_{0.5}$Si$_2$PO$_{12}$.

## Supplementary text 4

**AIMD simulation results for pristine LZSP**

We additionally compared the CHGNet MD results with AIMD on pristine LZSP even though DFT calculations are expensive for this supercell, to make sure the rationality of CHGNet MD simulations. To ensure sufficient migration events statistics, the simulation duration was lasted to 100 ps, with the temperature elevated to enhance ionic mobility, thereby achieving a sufficient large MSD of ~100 Å$^2$. This approach thereby ensured a good overlap between the two sets of curves, confirming the validity of the CHGNet MD data.

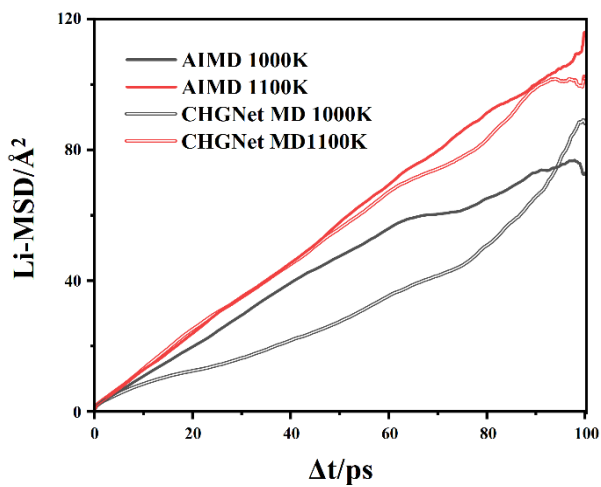

Fig S10. MSD curves of pristine LZSP calculated from AIMD and CHGNet MD simulations at 1000 K and 1100 K.

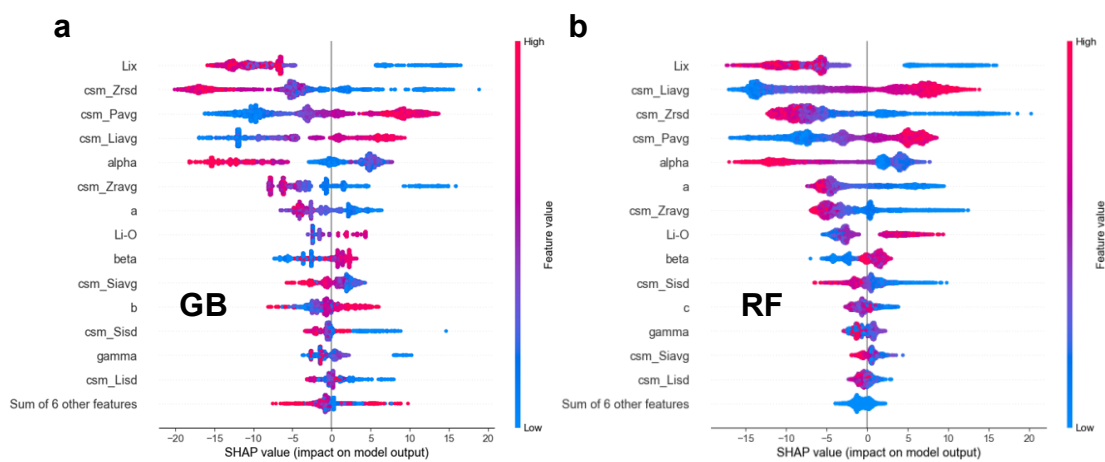

Fig S11. SHAP value for each feature in multinary (a) GB and (b) RF models, and points represent 4575 samples of 5-1 type in Table S2

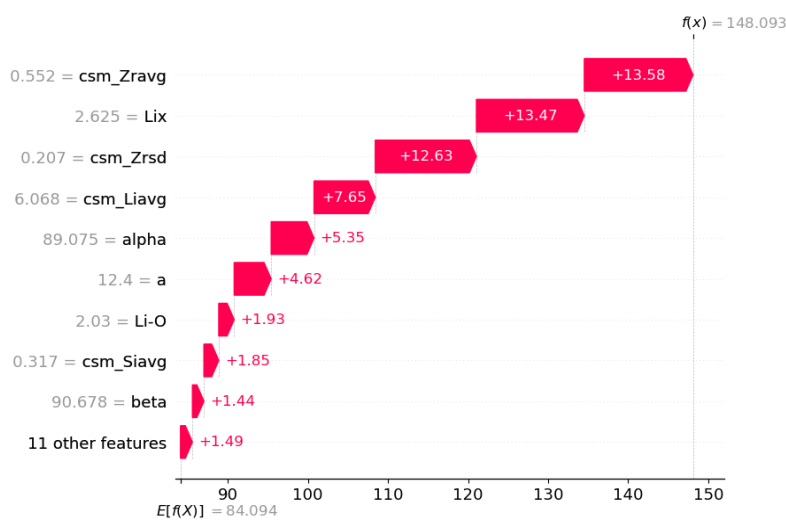

Fig S12. Waterfall plot for structure features of LZHSTNSP as predicted by multinary GB model.